\documentclass[12pt]{iopart}

\usepackage{graphicx,xcolor}
\usepackage{epstopdf}
\usepackage{dcolumn}
\usepackage{bm}
\usepackage{bbold}
\usepackage{float}
\usepackage[UKenglish]{babel}
\usepackage[utf8]{inputenc} 
\usepackage[centerlast,small]{caption}
\usepackage{subcaption}
\usepackage{lineno}
\usepackage[none]{hyphenat}

\usepackage{color}   
\usepackage{soul}    

\usepackage{hyperref}  
\usepackage[capitalise]{cleveref}   

\newcommand\myshade{85}
\colorlet{mylinkcolor}{black}
\colorlet{mycitecolor}{black}
\colorlet{myurlcolor}{black}

\hypersetup{
  linkcolor  = mylinkcolor!\myshade!black,
  citecolor  = mycitecolor!\myshade!black,
  urlcolor   = myurlcolor!\myshade!black,
  colorlinks = true,
}

\usepackage{siunitx}
\DeclareSIUnit\gauss{G}

\makeatletter
\renewcommand{\fnum@figure}{ Fig. \thefigure. }
\makeatother

\graphicspath{{Images/}}

\usepackage[backend=biber,
            citestyle=numeric-comp,
            style=phys,
            autocite=plain,
            maxbibnames=10,
            giveninits=true,
            sorting=none,
            isbn=false,url=false,eprint=false
            ]{biblatex} 
\renewbibmacro{in:}{}
\DeclareFieldFormat[article]{pages}{#1}
\AtEveryBibitem{\clearfield{month}}
\NewBibliographyString{available}
\DefineBibliographyStrings{english}{available = {},}
\DeclareFieldFormat{url}{\bibstring{available}\addcolon\space\url{#1}} 
\DeclareFieldFormat{urldate}{\addcomma\space\bibstring{urlseen}\space#1} 

\DeclareFieldFormat[article]{volume}{\textbf{#1}}
\DeclareNameAlias{sortname}{family-given}
\DeclareNameAlias{default}{family-given}

\DeclareFieldFormat{labelnumberwidth}{#1\adddot}

\begin{document}

\title[An Environmental Monitoring Network for Quantum Gas Experiments and Devices]{An Environmental Monitoring Network for Quantum Gas Experiments and Devices}
\author{T J Barrett$^1$, W Evans$^1$, A Gadge$^{1,2}$, S Bhumbra$^1$, S Sleegers$^1$, R Shah$^1$, J Fekete$^1$, F~Oru\v{c}evi\'{c}$^1$ and P~Kr\"{u}ger$^1$}
\address{$^1$ Department of Physics and Astronomy, University of Sussex, Brighton BN1 9QH, UK}
\address{$^2$ Department of Physics of Complex Systems, Weizmann Institute of Science, Rehovot 761001, Israel}
\ead{T.J.Barrett@sussex.ac.uk}


\begin{abstract}
Quantum technology is approaching a level of maturity, recently demonstrated in space-borne experiments and in-field measurements, which would allow for adoption by non-specialist users. Parallel advancements made in microprocessor-based electronics and database software can be combined to create robust, versatile and modular experimental monitoring systems. Here, we describe a monitoring network used across a number of cold atom laboratories with a shared laser system. The ability to diagnose malfunction, unexpected or unintended behaviour and passively collect data for key experimental parameters, such as vacuum chamber pressure, laser beam power, or resistances of important conductors, significantly reduces debugging time. This allows for efficient control over a number of experiments and remote control when access is limited.
\end{abstract}

\paragraph{\normalfont{\textbf{Introduction}}}
Recent developments in quantum technologies that exploit the unique properties of cold atomic clouds, such as gravimeters \cite{Menoret2018} and navigational accelerometers \cite{Cheiney2018, GarridoAlzar2019}, have been focused on producing miniature, field-based and remote systems. The challenging environmental conditions these systems are exposed to, as seen in space-borne experiments \cite{Becker2018}, can be mitigated using  automated control sequences, with evolutionary algorithms and machine learning protocols becoming increasingly common \cite{Wigley2016,Geisel2013,Tranter2018,Barker2020}. The rapid resolution of problems is essential in inter-dependent networks \cite{Canuel2020b} or in isolated remote systems where performance data may only be accessed infrequently, such as marine-based systems \cite{Bidel2018a, Zatezalo2008}.

Ultracold atom clouds are extremely sensitive to their environmental conditions due to having energy distributions on the nanokelvin scale. Typical laboratory-based systems experience drifts and instability caused by external environmental effects such as thermal disturbances and acoustic noise, which can affect a number of experimental components including mirrors, laser paths, coil resistances and power supplies used to generate magnetic fields. To mitigate these effects, it is possible to actively correct for individual issues with local feedback systems on parameters such as the polarisation of light beams \cite{Hidayat2008}, or the currents used to produce magnetic fields \cite{Thomas2020}, for example.

\begin{figure}[ht]
\centering
\includegraphics[width=\linewidth]{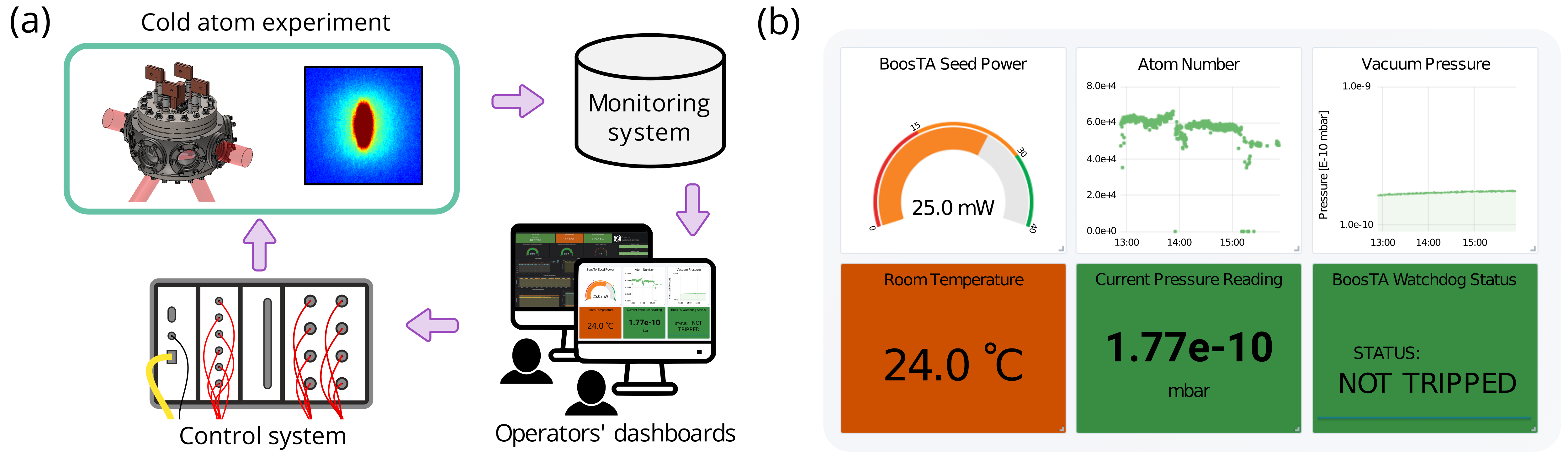}
\caption{(a) A diagram showing the systems that are needed for the monitoring, visualisation and control of an ultracold atom experiment. This demonstrates the role of the monitoring network in providing information to the user to improve experimental performance. (b) Example of a custom dashboard in a web browser interface used to assess performance of the experiments at any given time. Here, environmental data, such as room temperature and vacuum pressure are shown next to experimental results including atom number in real time.}
\label{fig:Overview}
\end{figure}

Often, environmental measurements are subsequently collected after an error has occurred, to retrospectively determine the underlying cause. This can be cumbersome and events may not be easy to reproduce, causing further operational delays. Here, we present a laboratory monitoring network used to autonomously record a number of experimental parameters across a suite of ultracold atom laboratories continually in the background, and visualise them in real time which allows for a fast response to unexpected errors. The ability to efficiently process and record environmental data will be crucial as quantum technology devices become more complex, for example using Bose-Einstein condensates (BECs) instead of thermal clouds, or as they are exposed to more challenging environmental conditions where changes are unpredictable. The described system employs a number of measurement nodes to record a wide variety of relevant parameters, such as temperatures, vacuum chamber pressures, laser power levels, and magnetic field strengths, as shown in Fig. \ref{fig:Overview} (a). The sensors are networked together and data are recorded in an external time-series database, which is then accessed through a series of end-user dashboards supported on an internet platform, an example of which is shown in Fig. \ref{fig:Overview} (b).

Our network contributes to an emerging ecosystem of laboratory monitoring systems that provide both measurement and management of environmental data \cite{Gurdita2016}. These database systems must be scalable, flexible and time-stamped for access in order to help users solve problems faster, and ideally remotely, facilitating experimental progress. The breadth of hardware and software tools that has been developed in the open source community means that monitoring systems can be designed for specific use, and implementation is straightforward. Improvements in the capability of programmable electronics \cite{Grinias2016, Kubinova2015, Zachariadou2012} and expanding database infrastructure can be exploited for use in cold atom laboratories \cite{Chilcott2021}. Microprocessor approaches have been implemented for control systems \cite{Malek2019, Eyler2011, Deng2012}, locking systems \cite{Eyler2013, Huang2014} and for environmental management \cite{Jaing2009}. Furthermore, such monitoring hardware can be integrated into feedback systems \cite{Mondal2013} and used with management software \cite{Jaing2009}. 

A similar advance has occurred in field programmable gate array (FPGA)-based hardware and has been followed by the development of FPGA-based control systems for atomic physics experiments \cite{BourdeauducqARTIQ, Perego2018}. Additional advances in control systems \cite{Keshet2013} have allowed for the automation of experimental feedback \cite{Starkey2013}, optimised control of complex sequences \cite{Amri2019} and even citizen-science participation \cite{Heck2018, Laustsen2021}. Remote control of laboratory experiments is common in situations such as particle physics \cite{Krause1997} and teaching laboratories to expand use of key equipment \cite{Hyder2010, Sharafutdinova2013}. Furthermore, as quantum technology approaches maturity, commercial systems are becoming available for remote non-specialist users through online portals \cite{ColdQuantaAlbert, IBMVigo}.

\begin{figure*}[ht]
\centering
\includegraphics[width=\linewidth]{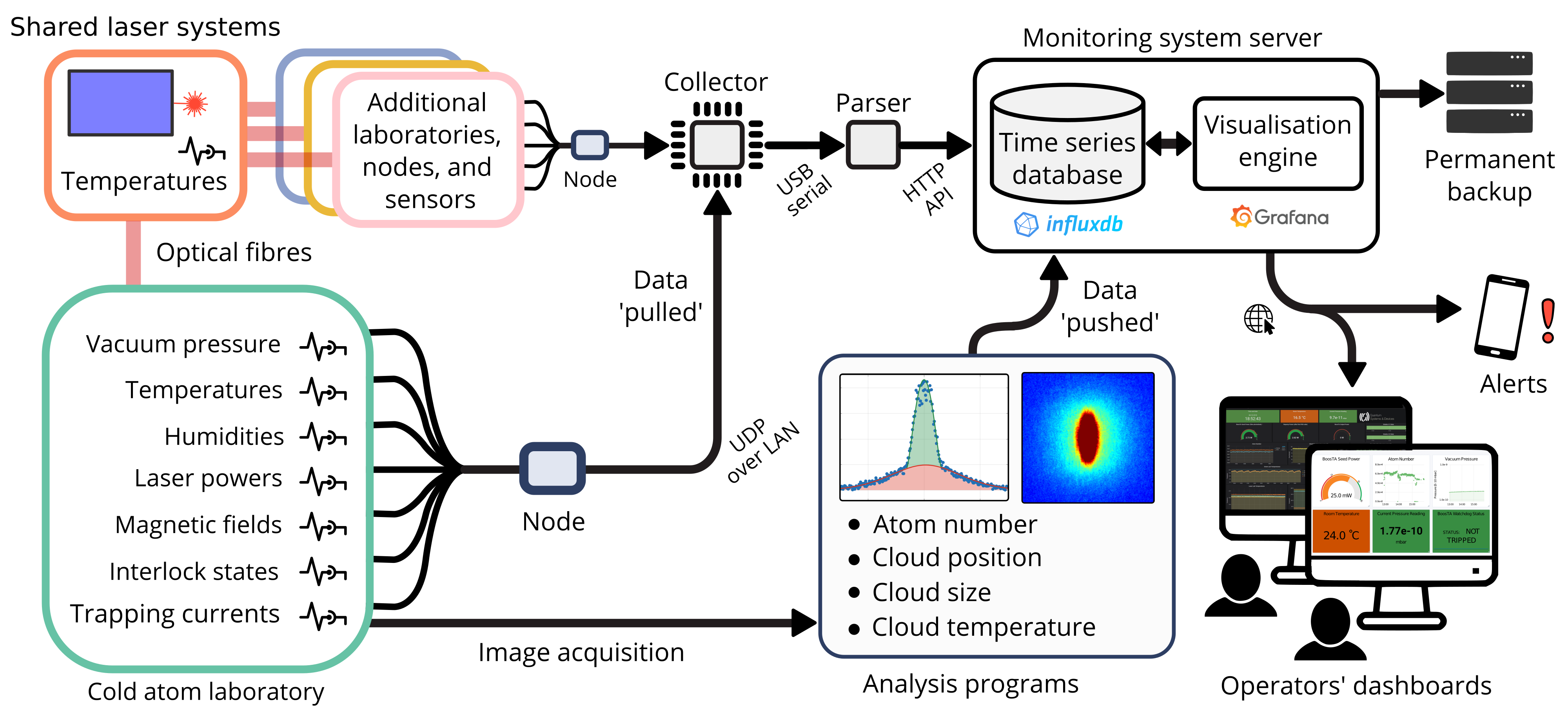}
\caption{A schematic showing the architecture of the monitoring system. On the left, environmental parameters are continually measured in both the local cold atom experiments and the shared laser system, and sent to the database via the system nodes and collector. Additional parameters are calculated through analysis of absorption images when they arrive, and are pushed directly to the database. The server hosts a time-series database, which stores all the data and periodically copies it to permanent backup storage. The stored measurements can be accessed through a visualisation engine, allowing for simple analysis, customised dashboards and alert systems for end-users.}
\label{fig:SystemArchitecture}
\end{figure*}
\paragraph{\normalfont{\textbf{Data acquisition}}}
A functional schematic of the entire monitoring network is depicted in Fig. \ref{fig:SystemArchitecture}. The system presented here has two types of measurement nodes: in the first type, data is `pushed' by the node to the database as part of the analysis protocol. As is typical in atomic physics experiments, this means the acquisition of a series of images of the atomic cloud \cite{Smith2011} to determine parameters such as atom number, cloud temperature, cloud shape, trap frequency, and more. Once these values are calculated, they are pushed directly to the database within the analysis programs. The second type of measurement nodes are microcontroller-based devices that locally acquire environmental measurements from various sensors around the laboratories when triggered via request over an isolated local area network (LAN), and as such data is `pulled' from these nodes in this case by the `collector' device. Organising the network in this way means that pulled environmental data is acquired in a synchronised manner with an update period set only by the collector device (in contrast to the atom cloud data, which is pushed whenever a new data point arrives). The vast array of microcontroller hardware available allows for each node to be configured to suit a range of requirements, and standardise data readings to conform to the database format. Current microprocessor hardware used here includes transimpedance amplifiers and photodiodes for measuring laser beam powers, Hall sensors for measuring magnetic field strengths, serial interfaces for communicating with devices such as vacuum gauges, and digital optocouplers for detecting the on/off states of equipment such as interlocks and shutters. The custom firmware written for all microcontrollers, schematics and designs for custom electronics, and other software used in this network has been made publicly available at an online repository \cite{Barrett2021}.

The existing ecosystem of microprocessor hardware is accessible, easy to implement, has community support available and elements can be integrated into custom devices to reduce the number of total components. For example, in the system presented here, a single printed circuit board (PCB) \cite{Barrett2021} was designed to consolidate 8 thermocouple amplifiers and 12 transimpedance amplifiers that are used to record the temperature at various locations and laser beam powers at several points along the beam paths, respectively. In this case, the thermocouple amplifier chip (MAX31855K, Maxim Integrated) converts the thermocouple readings to digital form and outputs them using the standard serial peripheral interface (SPI) communication protocol. Existing software communication libraries are available \cite{MAX31855Library} (as with all sensors used here), which makes reading a temperature into an array possible with a single simple command \verb|T[0] = thermocouple01.readCelsius()|. The availability of such software libraries facilitates fast setup of additional sensor nodes whenever required, for users with little programming experience.

The combination of different measurements is easily programmed and is carried out sequentially over each measurement node. Once a node receives a request for data from the collector device, it measures data from each of its attached sensors, collates them into a location-stamped string and returns this string via the LAN. There are over 100 sensors compiling environmental data across the devices in the system presented here. Messages are exchanged over a wired network to avoid wireless microwave noise interfering with the experiments - for instance, WiFi signals at \SI{2.4}{\giga \hertz} and \SI{5.0}{\giga \hertz} are close to the hyperfine ground states in alkali metals. The data are transferred via user datagram protocol (UDP), due to its simplicity and low-overhead, at intervals of \SI{20}{\second}, and was measured to have a $100\%$ transfer efficiency over a week of operation. This update rate was found to be appropriate for monitoring environmental data, but can be reduced to the millisecond level with the current microprocessor hardware (and the time-series database itself is limited to nanosecond time-stamps). Data is transferred in UDP packets, which are both sent and received easily using functions from an existing software library, \textit{EthernetUDP} \cite{ArduinoUDP}. For example, the collector device triggers a sensing node by sending a message simply by executing the commands  \verb|Udp.beginPacket(Node_IP, localPort); Udp.print("READ");| \verb|Udp.endPacket();| \cite{Barrett2021}. Finally, to preserve the robustness of communication, each device is programmed with a watchdog script to power-cycle reset should communication ever be lost, allowing them to automatically reconnect back to the network. 

\paragraph{\normalfont{\textbf{Database architecture}}}
Once the individual measurements have been taken at local nodes and sensors in each laboratory and transferred to the collector microcontroller over UDP LAN, as described in the previous section, it is important that they are then reliably imported to a time-series database (TSDB) for storage and management. The collector is connected directly to a `parser' device, based on a Raspberry Pi, using a USB serial communication bridge, which allows a single point of contact from the isolated LAN to the outside internet. The collector transfers any received measurement strings over to the parser, at which point Python programming scripts \cite{Barrett2021} are used to parse the data strings into separate variables, standardise the format, and enter them into a TSDB. The database is an instance of the open-source tool \textit{InfluxDB} \cite{InfluxDB}, which is running elsewhere on a network server with a backup storage drive. InfluxDB provides an application programming interface (API), and we use the Python library \textit{Requests} \cite{ReitzRequests} to easily post newly-parsed data to the hypertext transfer protocol (HTTP) endpoint of the API with commands of the form
\begin{verbatim}
requests.post("http://[IP]:[PORT]/write?db=[DB]", data=payload_string),
\end{verbatim}
where the user inputs the IP address, port number, and database name of the InfluxDB instance.

The TSDB software was chosen because it is optimised to efficiently write, compress, and store data, whilst also keeping it readily available for fast querying. This is important when maintaining historical data over years of use. The format is easily scalable and data points can be tagged and grouped for simple management. InfluxDB query language is simple, easy to integrate into existing analysis code, and similar to existing database languages such as structured query language (SQL). Each data entry is stored in the database with general the form:
\begin{verbatim}
"Measurement Name", <tag_key1>=<tag_value>,<tag_key2>=<tag_value> 
<field_key1>=<field_value>,<field_key2>=<field_value>
\end{verbatim}
with the \textit{tags} providing a way to add any relevant identifying labels to the measurements. Specifically, an example temperature measurement takes the following form:
\begin{verbatim}
"temperature", RoomID=Lab03, DevID=Dev01 T1=21.6, T2=22.8, T3=25.2.
\end{verbatim}
Each monitoring node in a network system can be adapted to suit the necessary local measurements, leading to a complex data set which we process and organise according to tags during collection at the parser. A system of ten measurement nodes, capturing an average of ten measurements each at 20 s intervals, requires  $\sim$ 6.25 GB of storage space per year. All data is automatically time-stamped on entry, is backed up regularly, and retention policies are used to downsample historic data ($>$ 1 year) from every 20 seconds to hourly values to conserve storage space.

Data visualisation tools are invaluable for interpretation of the stored data by general users not having detailed knowledge of the database scheme. Here, we use an open-source tool called \textit{Grafana} \cite{Grafana2019}, which has native support for direct integration with a TSDB. This combination of tools has been used for an expansive range of complex systems, from bee keeping \cite{HiveeyesProject} to financial monitoring \cite{GrafanaFinance}. We use the visualisation tool to allow users to build customised dashboards for displaying relevant metrics in a web browser from any remote location with an internet connection. This includes dashboards for different combinations of any measurements from several nodes across the network of laboratories - for example, in our setup we have a dashboard dedicated to displaying beam powers at several points along the chain of a shared laser system, over multiple rooms, to monitor stability at each point. Users can quickly look back at charts of data over any time period or see the overall system health at a glance rather than setting up additional time consuming measurements following an error.

For critical measurements the system is programmed to send an alert to users if a threshold is reached, for example if the temperature of a vacuum system rises too quickly during a bake out process which can damage components, or if the seed light injecting a tapered amplifier laser is too high or low. The ability to continuously assess a range of experimental parameters with one central database simplifies maintenance procedures which, if ignored, can otherwise cause long delays in experiments.

\paragraph{\normalfont{\textbf{Distributed laser system}}}
One example of the type of experimental infrastructure that can benefit from integration with a monitoring network is a collection of shared lasers. In our system, a pair of tapered amplifier lasers are used to provide the two central laser frequencies employed by a suite of three atomic physics laboratories, where this seed light is then locally amplified. Each laser is locked to an atomic reference in the central laboratory, split into individual paths for each experiment and then distributed to the laboratories via single-mode polarisation-maintaining optical fibers, as indicated in Fig.~\ref{fig:SystemArchitecture} (top-left corner). The ability to monitor properties of the light at several points along each path allows for fast debugging when necessary.

The optical power of the laser beams is monitored both in the centralised laser laboratory and in the individual science laboratories to assess the stability and detect changes in fibre coupling efficiencies. This is also important for the protection of the local amplifier chips which can be damaged if operated when the seed power is measured to be above or below a given threshold. Therefore, watchdog safety circuits, which switch off the amplifier if seed power deviates outside a preset range, have been implemented measuring both the input seed powers with  in-fibre photodiodes and amplifier output powers. Additionally, these circuits are integrated with interlock safety systems in each individual laboratory. The monitoring system is supplemented with local active feedback on the amplifier current, which is able to mitigate small drifts in the laser properties due to polarisation drift in the delivery fibres \cite{Kyselak2017, Eickhoff1981} and provide stable output power. This demonstrates the interaction between the monitoring system and local hardware to provide suitable protection and feedback. 

The ability to monitor laser powers at each point in the chain of this system, using a centralised dashboard, significantly reduces the time taken to rectify issues with coupling and frequency stability. This allows for simultaneous monitoring of a number of laboratories by a single user which paves the way for implementation of distributed networks of quantum technology devices, such as a network of gravimeters for gravitation-wave detection \cite{Badurina2020}.

\paragraph{\normalfont{\textbf{Integration with experimental data}}}
Current quantum technology devices have focused on using thermal atomic clouds but there are schemes which employ BECs for a greater sensitivity in gravimetry \cite{Aguilera2014, Dimopoulos2007} or a finer resolution in  microscopy \cite{Wildermuth2005, Gierling2011}. To generate BECs, we evaporatively cool a magnetically trapped cloud of atoms using radio-frequency (RF) radiation to a temperature below the critical value, typically on the order of merely hundreds of nanokelvin, which requires precise control of the atoms over an experimental cycle that is on the order of a few seconds to a minute. In addition to this, in order to perform long-term experiments with BECs and gather statistics, the system must run in a stable, repeatable way over at least a course of several hours. Such experiments are extremely sensitive to the external environment which, while making it challenging to produce BECs, provides substantial performance benefits for quantum sensors over their classical equivalents. 

Experimental systems designed for producing BECs involve complex parameter spaces. For example, just one laboratory in our system uses eight laser frequencies combined pairwise into eight beams, delivered to two different vacuum regions, using fifteen power supply currents, and three distinct stages of applied RF radiation. With our monitoring system recording many of these critical parameters in the background, the centralised database of measurements is easily accessed independently to download any specific time series. Additionally, since the experimental observables are also `pushed' to the database whenever they are calculated, as was indicated in Fig. \ref{fig:SystemArchitecture}, we are able to easily obtain the environmental measurements alongside the atom numbers, cloud temperatures, positions, and more, over any time period, allowing for simple debugging and stability studies of the entire system. The data can be downsampled and viewed on different timescales as needed to show both short-term and long-term variations.

\begin{figure}[ht]
\centering
\includegraphics[width=\linewidth]{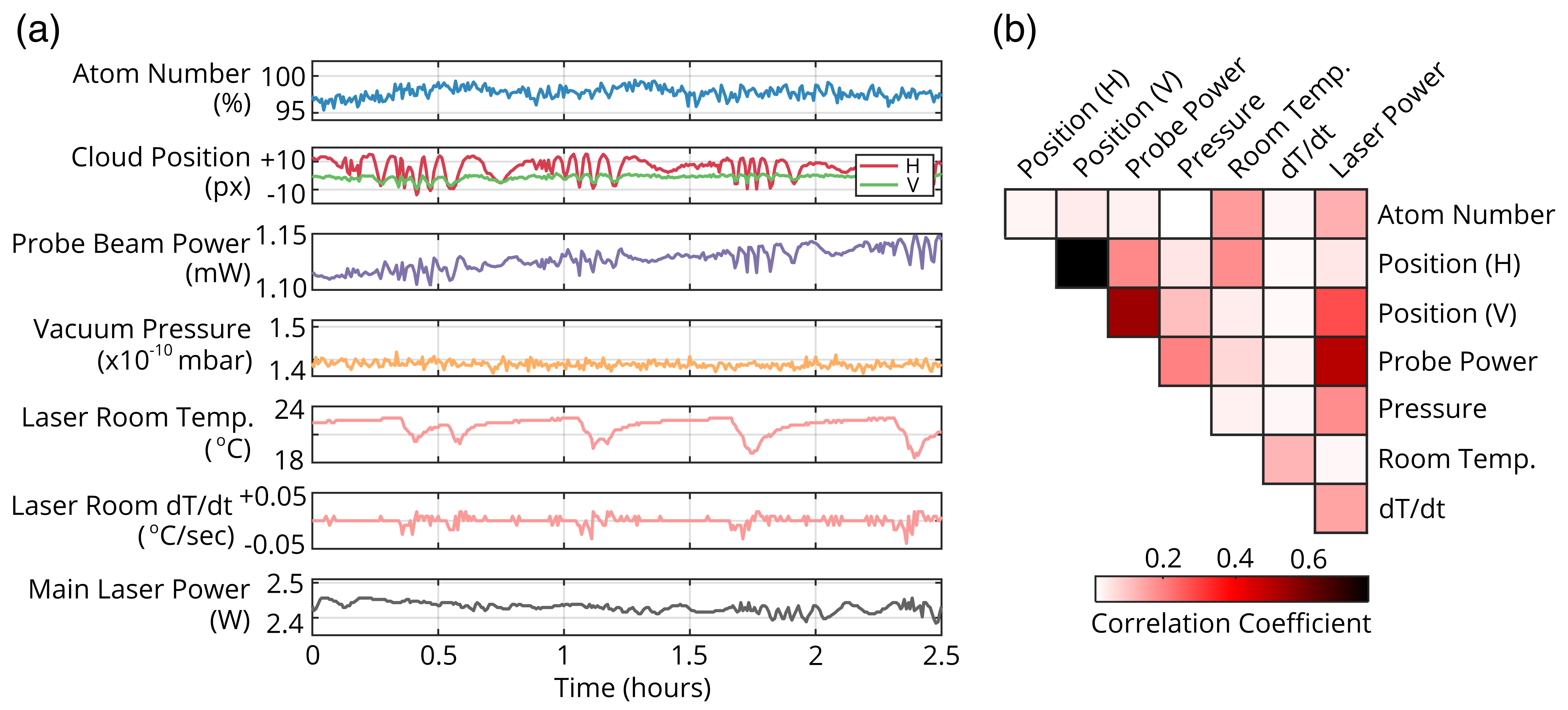}
\caption{(a) A collection of environmental measurements taken from the time-series database and experimental parameters calculated from image analysis during a repeated experimental run over two and a half hours. (b) An example correlation matrix calculated for the monitoring data above showing Pearson coefficients. There are strong correlations between the cloud position in both horizontal (H) and vertical (V) directions with the power in the absorption imaging probe beam. However, there are only strong correlations between the cloud V position and the output power of the local laser amplifier, showing that the cloud position is more sensitive to the laser power in this direction. The strongest correlations with atom number are seen in the local laser output power and central laser laboratory temperature, which highlight problems due to fluctuations in seed power introduced during fibre coupling.}
\label{fig:Correlations}
\end{figure}

An example showing the collated raw monitoring system data for a magneto-optical trap (MOT) stage --- one of the pre-cooling stages in a BEC experiment, when the atoms are still at a temperature of several hundred microkelvin --- is presented in Fig.\;\ref{fig:Correlations}~(a). There are a variety of signal processing techniques which can be applied to characterise the correlations between the captured signals. As an example, in Fig.\;\ref{fig:Correlations}~(b) we characterise the linear dependence between the variables by constructing a correlation matrix of Pearson coefficients, which is a scaled covariance, as this technique highlights large correlations that can point to specific experimental problems. For example, both the horizontal (H) and vertical (V) cloud positions are strongly correlated with the power in the absorption imaging beam used to probe the atomic sample (Probe Power), which in turn is derived from the main laser. The atom number has its largest correlations with local laser output power and temperature in the central laser laboratory, which was traced back to sub-optimal optical fibre coupling alignment and associated polarisation drifts.

At this point, with access to such data, further techniques could be used for identifying a more abstract analysis that best describes the entire system. Here however, we maintained the original variables throughout for simplicity, as it was found to be sufficient for the inspection and debugging process. Nevertheless, for example, time-lagged linear dependencies would be uncovered by using cross-correlations to find phase differences. Similarly, frequency responses and correlations in frequency space could be determined by comparing each signal's power spectral density from a discrete Fourier analysis. Furthermore, the analysis of such a wide data set can be extended in detail using machine learning techniques and principal component analysis \cite{Segal2010}.

\begin{figure}[ht]
\centering
\includegraphics[width=\linewidth]{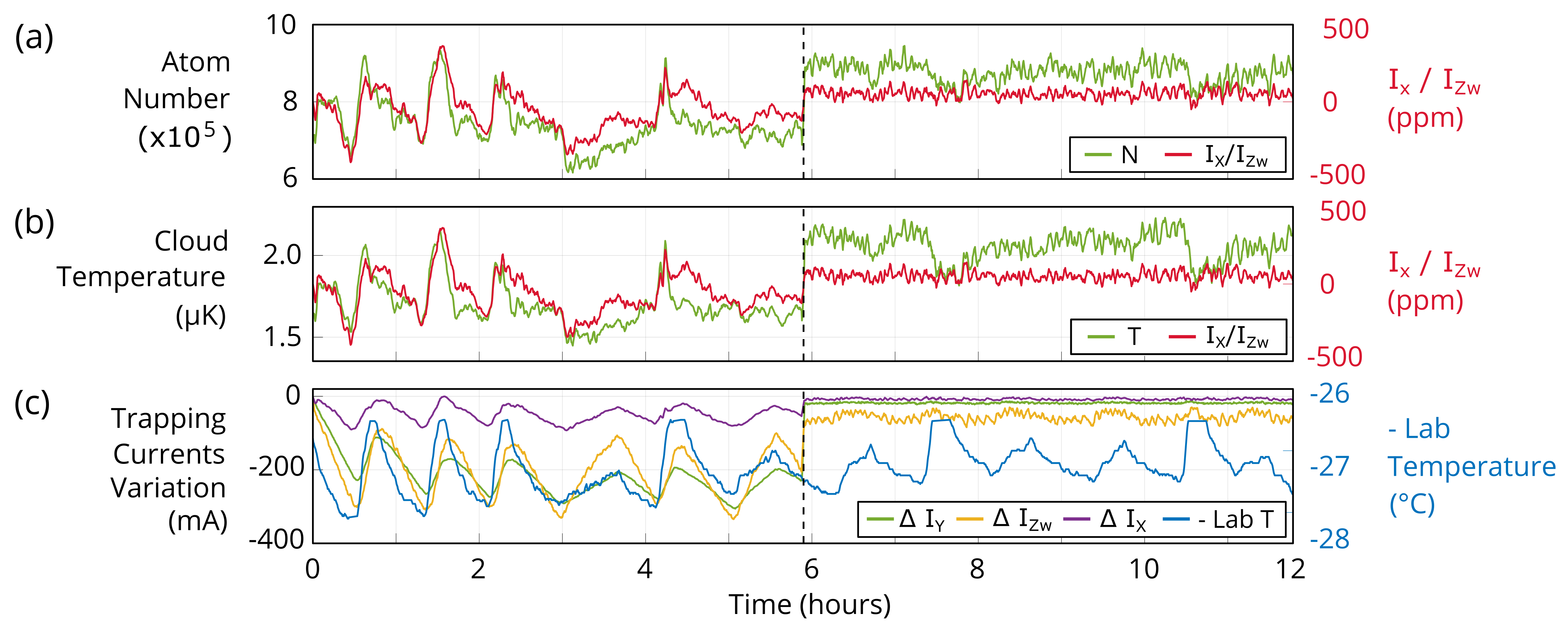}
\caption{Stability data collected by the monitoring system for a \SI{1.6}{\micro\kelvin} cloud, just above the transition to BEC. The variation in atom number (a) and cloud temperature (b) are shown together with the ratio of trapping currents I$_\textrm{X}$/I$_\textrm{Zw}$, exhibiting strong correlation with both. (c) The variation of the three individual electric currents I$_\textrm{Zw}$, I$_\textrm{X}$, and I$_\textrm{Y}$ (driven through the Z-wire, and X-bias and Y-bias field coils, respectively) used to generate the trapping potential are plotted together with the ambient laboratory temperature. Note that the negative of the temperature is displayed, to emphasise the anti-correlation with the currents (i.e., increase in temperature leads to increase in conductor resistance, and therefore a decrease in current flow for the same voltage). After the first six-hour time period (indicated by the vertical dashed line), a current feedback system was activated, leading to a significant improvement in the stability of the cloud temperature and atom number.}
\label{fig:Stability_currents}
\end{figure}

We now turn to the example of a much colder cloud, just above the phase transition temperature to BEC, which is a system particularly sensitive to variations in the external environment. Examples of aggregated data from the monitoring system for such a system are shown in Fig. \ref{fig:Stability_currents}. In this measurement, $^{87}$Rb atoms are spin-polarized in the $|F=2,m_F=+2\rangle$ magnetic sub-state and confined in an harmonic Ioffe-Pritchard style trap, which is created by passing electrical current through a Z-shaped wire together with two external homogeneous bias fields (denoted X-bias and Y-bias) \cite{reichel1999atomic}. The atoms are then prepared at a temperature of around \SI{1.6}{\micro\kelvin} by an RF evaporative cooling sweep down to a fixed frequency, thereby truncating the trap depth \cite{davis1995evaporative}. The most critical properties of the cloud, namely atom number and temperature, are extracted through optical absorption imaging with a resonant probe beam \cite{Smith2011}, and the procedure was repeated every 20 seconds with a new atomic sample in order to examine the stability of the experimental system.

It can be seen in Fig. \ref{fig:Stability_currents} (a) and (b) that the cloud exhibits significant instability and drifts over the first six hours. Specifically, a peak-to-peak variation of 40\% of the total atom number, and \SI{0.7}{\micro\kelvin} in the cloud temperature are observed. Such variations of the important observables are known to be a common problem for experimental ultracold systems, and it is crucial to track down and mitigate those instabilities at their source. At this point, the action taken would typically be to begin collecting data to identify the contributing quantities, which requires significant additional time and resources. It is in exactly this type of situation where an environmental monitoring system becomes an extremely useful tool. In this case, since the system is continually collecting data from a variety of sensors we were able to simply retrieve a range of measurements from the relevant time period, to uncover the source of the problem. 

Figure \ref{fig:Stability_currents} (c) shows such collected data from measurements of the electrical currents used to drive the Z-shaped wire (I$_\textrm{Zw}$) and the two coils generating the orthogonal homogeneous bias fields (I$_\textrm{X}$ and I$_\textrm{Y}$), which combine together to create the trapping potential for the atom cloud. The currents are continually measured with a series of fluxgate-based electrical current transducers (LEM CASR 50-NP).  In contrast to the atom number drifts, a variation of up to several hundred milliamps that is more periodic in nature is apparent for all three currents, and no one individual current appears obviously responsible. The periodic variation is in turn identified to be caused by the ambient laboratory temperature [also plotted in Fig. \ref{fig:Stability_currents} (c)]. However, the \textit{ratio} of I$_\textrm{X}$/I$_\textrm{Zw}$ has been plotted in addition on Fig. \ref{fig:Stability_currents} (a) and (b), and is well-correlated with both the atom number and temperature. This can be understood because, by the construction of the trapping configuration, the ratio of the two currents I$_\textrm{X}$ and I$_\textrm{Zw}$ is in fact responsible for prescribing the value of the magnetic field at the trap bottom. Therefore, since the absolute final RF cooling frequency is fixed, a trap bottom change results in a trap depth change, and ultimately to a loss of higher energy atoms and an associated reduction in cloud temperature. In contrast, the remaining current I$_\textrm{Y}$ primarily controls the position of the trap minimum, and has only a very weak effect on the field at the trap bottom.

After identifying the largest contribution to the cloud variations, we implemented a simple feedback system for the electrical currents by combining information of the measured currents with the voltages applied across the loads to obtain the instantaneous resistances of each conductor, and adjust the power supply voltages as necessary to minimise variations in currents for each experimental cycle. The results after activating the feedback can be seen during the second six-hour period of Fig. \ref{fig:Stability_currents}, showing that the peak-to-peak variations have been reduced to 15\% of the total for the atom number (a factor of 2.7 improvement), and to \SI{0.35}{\micro\kelvin} for the cloud temperature (a factor of 2 improvement), in spite of the ambient laboratory temperature continuing to vary as before. This workflow demonstrates that the ability to concurrently measure multiple environmental parameters in tandem with experimental observables allows for targeted improvements of experimental performance without the need for a large series of independent measurements retrospectively.

\begin{figure}[ht]
\centering
\includegraphics[width=\linewidth]{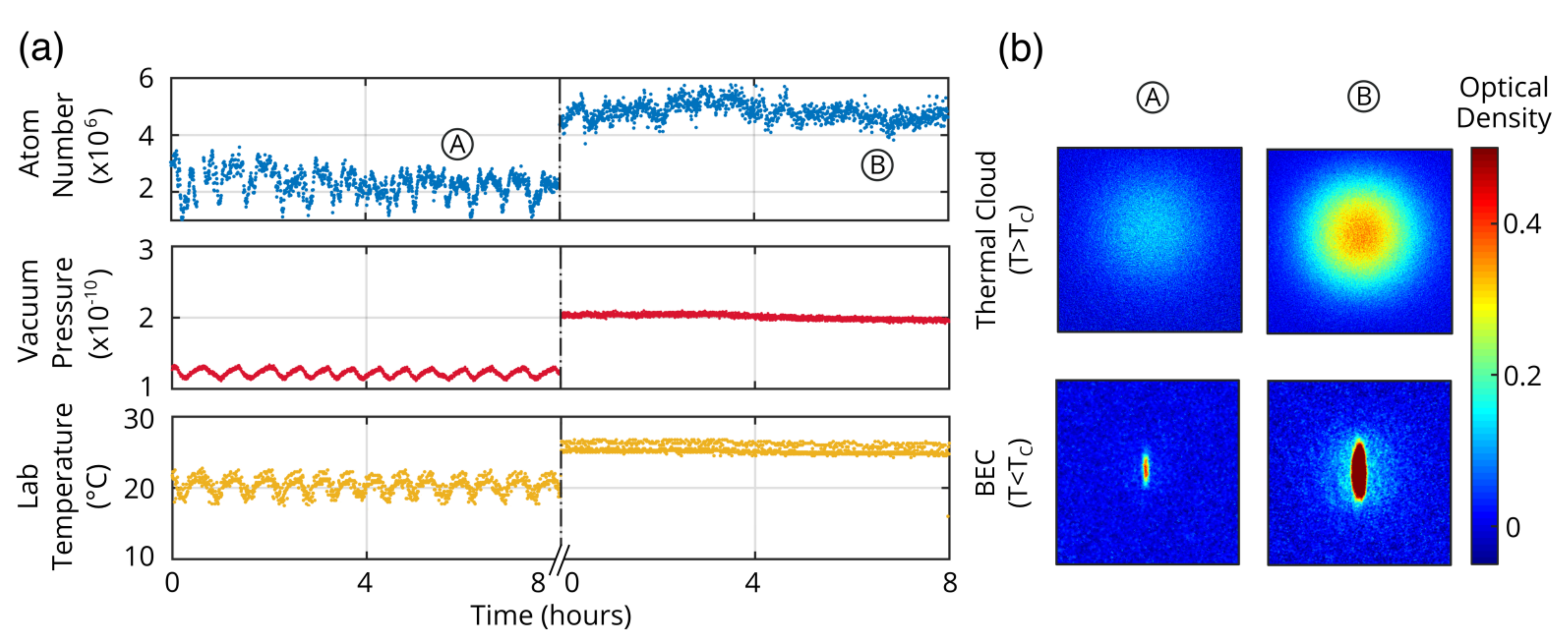}
\caption{Comparison of the atom number in a thermal cloud and BEC, in a magnetic trap under different environmental conditions. (a) The atom number of a magnetically-confined thermal cloud in one experiment in the network is plotted with a measurement of the vacuum chamber pressure and the ambient room temperature. After the first eight-hour period (indicated by the vertical dashed line) the local air-conditioning system is deactivated. The laboratory temperature changes from $(20.1 \pm 1.2)$ \SI{}{\degreeCelsius} to $(25.1 \pm 0.3)$ \SI{}{\degreeCelsius}, and the vacuum pressure from $(1.22 \pm 0.05) \times 10^{-10}$ \SI{}{\milli\bar} to $(2.01 \pm 0.04) \times 10^{-10}$ \SI{}{\milli\bar}. The atom number correspondingly increases from $(2.2 \pm 0.6) \times 10^{6}$ to $(4.8 \pm 0.3) \times 10^{6}$. (b) A pair of absorption images of atomic clouds is shown from each eight-hour window, at times indicated by the labels A and B. The images for each of the two times depict both a trapped thermal cloud (above the transition temperature, $T_c$), and the corresponding BEC that is produced by further evaporative cooling in each case.}
\label{fig:Stability_pressure}
\end{figure}
Furthermore, the data captured with the monitoring system allows for continuous comparison of the experiment under different environmental conditions. The example in Fig. \ref{fig:Stability_pressure} shows the increase of the atom number in a BEC, permitted due to better optimisation and more stable environmental parameters. After the first eight-hour period of collecting measurements, the local air-conditioning system was deactivated, to investigate its effect. It can be seen firstly that the overall ambient temperature increased, and secondly that the saw-tooth behaviour in the temperature profile --- characteristic of bang-bang style (on/off) feedback controllers in commercial air conditioning systems \cite{Al-Azba2020} --- has been eliminated. The room temperature in turn affects the vacuum chamber pressure, and the overall effect is an increase in the atom number (along with a reduction in variation) in the magnetic trap just above the transition temperature from $(2.2 \pm 0.6) \times 10^{6}$ to $(4.8 \pm 0.3) \times 10^{6}$. This ultimately results in a much larger atom number in the final BEC, illustrated in Fig. \ref{fig:Stability_pressure} (b). 

These examples show the benefit of having access to regular environmental data to improve experimental performance by optimising the system to the most stable background conditions. The ability to autonomously monitor a series of relevant variables, including power supply currents, conductors' resistances, and laser powers, is key to allow experimental development even with limited laboratory access for debugging, as for in-field devices or up-scaled quantum technologies. Using this system allowed one of our laboratories to advance an experiment from using thermal clouds to producing BECs without physical access to the laboratory during the Covid-19 pandemic in 2020.

\paragraph{\normalfont{\textbf{Conclusion}}} 
The accessible and flexible monitoring system presented here provides a useful blueprint for replication in similar cold atom laboratories. The ability to continually measure key experimental parameters and centralise data storage across a network of experiments paves the way for consolidated control and shared infrastructure, demonstrated by the distributed laser system. The time-stamped database system allows a wide range of measurements from various sources to be aggregated into one place in a standardised format and presented to end users with versatile dashboards, enabling the general health of the experiments to be assessed at a glance. This reduces the time needed for problem solving from days or hours down to minutes or seconds, and eases the progression of experiments, shown here by the remote creation of a Bose-Einstein condensate in an experiment which did not previously have one. We have demonstrated the usefulness of such a monitoring system in identifying the sources of long-term instabilities for both a magneto-optical trap and a magnetically-trapped ultracold atomic cloud just above the phase transition temperature to BEC, as well as the benefits emerging from the ability to monitor and optimise a BEC itself. The matrix of correlation coefficients illustrates how the real-time analysis of experimental and environmental parameters can highlight seemingly unintuitive dependencies. The integration of such analytical tools is essential in democratising quantum technologies as they assist the end-users without specialist knowledge in quantum physics in operating and maintaining complex systems. The data management infrastructure presented here can be employed in a wide range of quantum setups, including large networks of atomic experiments or remote systems where access is limited, in a laboratory context or in the field.

\section*{Data availability}
Further details required to reproduce the system described here have been made publicly available at an online repository \cite{Barrett2021}. This includes: datasheets of commercial devices integrated into our system; schematic and PCB production drawings (Gerber and NC drill files) for bespoke electronic circuits; microprocessors’ firmware; parser python scripts; and a collection of user guides and reports written by authors detailing the implementation and integration of individual system components.

\nolinenumbers
\section*{References}
\printbibliography[heading=none]

\typeout{get arXiv to do 4 passes: Label(s) may have changed. Rerun}
\end{document}